\documentclass[twocolumn,pra,aps]{revtex4}

\usepackage{graphicx}
\usepackage{amsmath}
\usepackage{latexsym}
\usepackage{amsfonts}
\usepackage{amssymb}
\usepackage{array}
\usepackage{epsfig}

\begin{document}

\title
{Decoherence of Highly Mixed Macroscopic Quantum Superpositions}

\author{Hyunseok Jeong$^{1,2}$, Jinhyoung Lee$^3$ and Hyunchul Nha$^4$}

\address{
$^1$Centre for Quantum Computer Technology,
Department of Physics, University of Queensland, Brisbane, Qld 4072,
   Australia\\
$^2$Department of Physics and Astronomy, Seoul National University, Seoul 151-747, Korea \\
$^3$Department of Physics, Hanyang University, Seoul, Republic of Korea\\
$^4$Department of Physics, Texas A \& M University at Qatar, Doha, Qatar}

\date{\today}

\begin{abstract}
It is known that a macroscopic quantum superposition (MQS), when it is exposed to environment, 
decoheres at a rate scaling with the separation of its component states in phase space.
This is more or less consistent with the well known proposition that a more macroscopic
quantum state is reduced more quickly to a classical state in general.
Effects of initial mixedness, however, on the subsequent decoherence of MQSs have been less known.
In this paper, we study the evolution of a highly mixed MQS interacting with an environment, 
and compare it with that of a pure MQS having the same size of the central distance between its %%@
component states.
Although the decoherence develops more rapidly for the mixed MQS in short times, its rate can be %%@
significantly suppressed 
after a certain time and becomes smaller than the decoherence rate of its corresponding pure MQS. 
In an optics experiment to generate a MQS, our result has a practical implication that
nonclassicality of a MQS can be still observable in moderate times even though a large amount of %%@
noise is added to the 
initial state.
\end{abstract}

%\pacs{PACS number(s); }

\maketitle

%Uncomment for PACS numbers title message
%\pacs{00.00, 20.00, 42.10}
% Keywords required only for MST, PB, PMB, PM, JOA, JOB? 
%\vspace{2pc}
%\noindent{\it Keywords}: Article preparation, IOP journals
% Uncomment for Submitted to journal title message
%\submitto{\JPA}
% Comment out if separate title page not required

\section{Introduction}

The behaviors of microscopic quantum systems are radically different from our
everyday experience in the macroscopic world.
The superposition principle of quantum mechanics plays the crucial
role for counter-intuitive behaviors of quantum objects. 
Since Shr\"odinger's famous illustration
of a macroscopic object in a quantum superposition
[1], there have been great interests 
in manipulating and observing macroscopic quantum superpositions (MQSs)
 [2-6]. However, decoherence is known as the main obstacle
to this attempt [7]. It is well known that a MQS
loses its quantum properties, through its interaction with environment, much faster than a %%@
microscopic one [7,8]. 
In this context, decoherence is often used to explain how the classical world appears
from the microscopic entities which individually obey quantum mechanical principles [7].

A superposition of two coherent states that are distinctly separated in phase space is
a well known example of a MQS [9,10]. This pure MQS shows nonclassical properties
such as negativity of the Wigner function and interference fringes in phase space [10].
Remarkably, it has been proved that separation between the component states of
a MQS is a crucial factor determining its decoherence rate [11-13].
The more separated in the phase space the component coherent states are, the faster the quantum %%@
features disappear.
A similar trend is also confirmed in the decoherence of Einstein-Podolsky-Rosen (EPR) correlations %%@
[14].
It is more or less consistent with the well known observation that 
more macroscopic quantum superpositions decohere faster in general
[7]. 
In practical situations, the quantum system is not always in a pure state initially, 
but can be in a certain mixed state by various experimental imperfections. However, effects of %%@
initial mixedness 
on the subsequent decoherence of MQSs have not been studied in detail even though truly macroscopic %%@
physical systems are 
typically in mixed states.

Recently, Jeong and Ralph showed that a {\it highly mixed} MQS can also exhibit strong quantum %%@
properties even 
as its entropy (i.e. mixedness) becomes extremely large [15].
One such example is a superposition
of thermal states separated in the phase space [15].
This result raises another related, interesting, questions concerning
decoherence of MQSs: Which state is more robust against decoherence between a highly mixed MQS and a %%@
pure MQS
when their component states are equally separated in the phase space?
Also, which state is more robust against decoherence
when they are equally macroscopic, i.e., 
the sizes of the physical systems are equal?
More generally, how could mixedness affect the evolution of decoherence in MQSs?

In this paper, we show that nonclassicality of
a highly mixed MQS can disappear more slowly after a certain time than that of a pure MQS when 
the centers of their component states are equally separated.
This implies that initial mixedness may not be very detrimental to the observation of nonclassical %%@
properties for 
a MQS in moderate times. For example, in an optics experiment to generate a MQS,
nonclassicality of a MQS can be still observable even though
the Gaussian noise is inevitably added to the initial state.
We also make a similar comparison between
a pure MQS and a mixed MQS that are ``equally macroscopic" in the sense that their
average photon numbers are the same. In this case, nonclassicality of a mixed MQS disappears %%@
conspicuously slower
than that of a pure MQS. 
These two observations are explained by the fact that the initial mixed MQS can be decomposed into a %%@
sum of pure MQS's, 
some of which are more robust against decoherence than the single pure MQS. 
Our results provide a practical basis for observing
fragile macroscopic quantum phenomena, and imply that the total energy quanta may not be a good %%@
indicator of decoherence 
rate beyond short-time regime.

\section{Pure and mixed macroscopic quantum superpositions}

We first introduce
the pure MQS [9,10].
A coherent state, $|\alpha\rangle$,
when its amplitude $\alpha$ is large, is known as most classical
among all pure states [16].
A superposition of two coherent states (SCS),
\begin{equation}
\label{PMQS}
|\Psi_{\alpha}\rangle=
{\cal N}_\alpha
(|\alpha\rangle-|-\alpha \rangle),
\end{equation}
where ${\cal N}_\alpha^{-1}=\sqrt{2-2e^{-2|\alpha|^2}}$,
is considered a MQS for $|\alpha|\gg1$ [17].
Very recently, such SCSs were %performed
%in cavity fields and 
experimentally generated
in free-traveling fields [18].
We stress that the SCS in Eq.~(\ref{PMQS}) may be considered
a MQS only if the amplitude $\alpha$ is sufficiently
large. In this regime,
the average photon number of the state is very large and
the two coherent states, 
$|\alpha\rangle$ and $|-\alpha\rangle$, are macroscopically distinguished. 
We shall suppose this condition 
throughout the paper. 
% $\phi=\pi$ throughout this paper.

Now, let us consider a different type of MQS with initial mixedness as follows.
A coherent state of amplitude $\alpha$, when exposed to a Gaussian noise with variance $V$, is %%@
represented by
\begin{equation}
\rho^{th}(V,\alpha)=\int d^2\beta P_\beta^{th}(V,\alpha)
|\beta\rangle\langle\beta|
\label{DTS}
\end{equation}
where
\begin{equation}
P_\beta^{th}(V,\alpha)=\frac{2}{\pi(V-1)}
\exp[-\frac{2|\beta-\alpha|^2}{V-1}].
\end{equation}
A mixed MQS with sufficiently large $\alpha$
can be represented as
\begin{equation}
\label{MMQS}
\rho= {\cal N}
\Big(\rho^{th}(V,\alpha)
+\rho^{th}(V,-\alpha)
-\sigma(V,\alpha)\Big),
\end{equation}
where 
$\sigma(V,\alpha)=\int d^2\beta P^{th}(V,\alpha)|\beta\rangle\langle-\beta|+H.C.$
and ${\cal N}=(2-2\exp[-2\alpha^2/V]/V)^{-1}$.
When $V=1$, the state $\rho$ becomes a pure MQS in Eq.~(\ref{PMQS}).
It was shown that this mixed MQS can be generated when
a displaced Guassian state in Eq.~(\ref{DTS}) is used as the input state,
instead of a pure coherent state, to the SCS 
generation process [15].
The mixed state (\ref{MMQS}) shows
strong nonclassical properties 
regardless of the values of $\alpha$ or $V$ [15].
Note that quantum behaviors of the MQS are due to
the coherence term $\sigma(V,\alpha)$.
If $\sigma(V,\alpha)$ were zero,
the mixed state (\ref{MMQS}) would become
a mere classical mixture of the two local Gaussian states
without any quantum properties.

Mixedness of a state $\rho$ can be quantified
by its linear entropy $S(\rho)=1-\mathrm{Tr} (\rho^2)$. 
The degree of mixedness of the state in Eq.~(\ref{MMQS}) is given by
\begin{equation}
S(\rho)=1-
4{({\cal N})}^2
\Big(\frac{1+\exp[-\frac{\alpha^2}{V}]}{V}
-\frac{4\exp[-\frac{4\alpha^2V}{1+V^2}]}{1+V^2}
\Big).
\end{equation}
We shall say that state $\rho$ is {\it highly mixed} 
when $S(\rho)>0.99$. 

The Wigner function of a quantum state is 
a quantum mechanical analogy of the probability distribution in phase space
[19,20], and it can take negative values unlike classical probability distributions. 
This negativity is regarded as a clear signature of nonclassicality of a physical system
[21].
In order to experimentally observe 
negative values of the Wigner function, which we shall often call simply ``negativity" in this %%@
paper,
the size of negativity must be large enough.
This is particularly true in experiments with limited detection efficiency.
For example, homodyne detection in quantum optics can be used to
reconstruct the Wigner function. However,
the efficiency of homodyne detection cannot be perfect in real experiments and
small negative values of Wigner functions are hard to directly observe.

\section{Decoherence}

It was pointed out that the decoherence rate of pure MQS scales with 
the distance between the component states [11,12].
A similar trend is also confirmed in 
the decoherence of quantum
nonlocality for EPR states in the thermal environment [14]. 
In Ref.[14], Jeong {\it et al.} showed that the more strongly 
the initial field is squeezed (i.e. closer to the ideal EPR state),
the more rapidly the maximum nonlocality decreases. 
Note that in the limit of the ideal EPR state, the average photon
number at each mode approaches infinity. 
In particular, the authors of Ref.[14]
explained the scaling of vanishing nonlocality with the degree of squeezing as follows. 
An EPR state can be understood as
a multi-mode superposition of two-mode coherent states. 
As the degree of squeezing
becomes larger, the superposition between component coherent states
extends further so that the average separation between the component
states becomes larger.
This causes the quantum coherence, more precisely, quantum nonlocality, to be
destroyed more rapidly.

In this paper, we compare the decoherence of the mixed MQS (actual state) in Eq.~(\ref{MMQS}) 
with that of the pure MQS (target state) in Eq.~(\ref{PMQS}) for the same values of $\alpha$, 
namely, under the condition that the central distances between their component states are equal. 
In addition, we also look into the evolutions of the two MQSs, one pure and the other mixed, at the %%@
same levels of ``macroscopicity".
For this purpose, the average photon number of MQS is adopted
as the size of macroscopicity.
The average photon number ${\bar n}$ of 
the mixed MQS (\ref{MMQS}) is
\begin{equation}
{\bar n}={\rm Tr}[\hat a^\dagger \hat a \rho]=
 \frac{{\cal N}\exp[\frac{2\alpha^2}{V-1}]
 \{V(V-1)Q_{(5)}+2\alpha^2Q_{(7)}\}}{V^3(V-1)}
\end{equation}
where  $\hat a^\dagger$ ($\hat a$) is the creation (annihilation) operator and 
$Q_{(n)}=\exp[-\frac{2\alpha^2}{V-1}]V^{(n/2)}(V-1)\sqrt{V+V^{-1}-2}$.
Note that the average photon number of the pure MQS
is ${\bar n}=\alpha^2\coth[\alpha^2]$.

\begin{figure}
\centerline{\scalebox{0.75}{\includegraphics{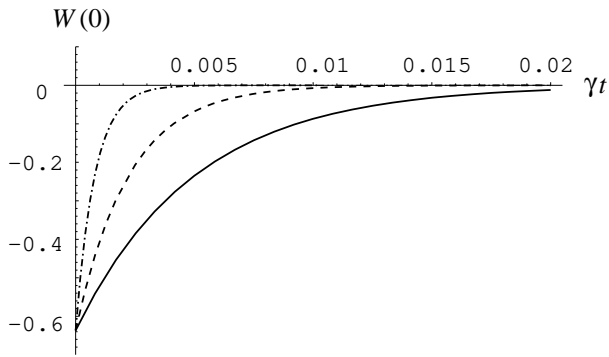}}}
\caption{
The minimum negativity of the Wigner functions,
$W(0)$, of pure MQSs for $\alpha=20$
(solid curve), $\alpha=30$ (dashed curve) and $\alpha=50$ (dot-dashed curve)
against the scaled time $\gamma t$. 
The minimum negativity of a MQS approaches zero more rapidly when $\alpha$ is larger.}
\label{pure}
\end{figure}

The decoherence of the MQSs
can be studied by solving the master equation [22]
\begin{equation}
\label{master-eq}
  \frac{\partial \rho}{\partial t}=\hat{J}\rho +\hat{L}\rho~;~\hat{J}
\rho=\gamma a\rho a^\dag,~~
  \hat{L}\rho=-\frac{\gamma}{2}(a^\dag a\rho +\rho a^\dag a)
\end{equation}
where $\gamma$ is the energy decay rate
and $t$ is time. In the above equation, 
we assume the optical-frequency regime in which the average photon number in the reservoir 
is negligible even at room temperature. The well known solution of the master equation
for a coherent-state dyadic $|\alpha\rangle\langle\beta|$ can
be described as
\begin{equation}
|\alpha\rangle\langle\beta|\rightarrow
e^{-(1-\kappa)\{(|\alpha|^2+|\beta|^2)/2
-\alpha\beta^*\}}
|\sqrt{\kappa} \alpha \rangle\langle \sqrt{\kappa}\beta |,
\end{equation}
where $\kappa\equiv e^{-\gamma t}$.
Using this solution,
the superposition of thermal states after time $t$ is
given by 
\begin{equation}
\label{MMQS2}
\rho(t)= {\cal N}(t)
\Big(\rho^{th}(V^\prime,\alpha^\prime)
+\rho^{th}(V^\prime,-\alpha^\prime)
-\sigma^{\cal C}(V^\prime,\alpha^\prime)\Big),
\end{equation}
where
\begin{equation}
\sigma^{\cal C}(V^\prime,\alpha^\prime)=\int d^2\beta
P^{th}(V^\prime,\alpha^\prime)e^{-2(1-\kappa)|\beta|^2}
|\beta\rangle\langle-\beta|+{\rm H.C.},
\end{equation}
$\alpha^\prime=\sqrt{\kappa}\alpha$,
$V^\prime=\kappa(V-1)+1$, and
\begin{equation}
{\cal N}(t)=2-\frac
{8\exp[-\frac{2\alpha^2(3\kappa+1)}
{3\kappa(V-1)+(V+3)}]}
{3\kappa(V-1)+(V+3)}.
\end{equation}

The Wigner function $W(\eta)$ of a density operator $\rho$
can be obtained as
\begin{equation}
W(\eta)
=\frac{1}{\pi^2}
\int d^2\xi
e^{\eta\xi^*-\eta^*\xi}\chi(\xi)
\end{equation}
where $\chi(\xi)$ is
the Weyl characteristic function
$\chi(\xi)=\mbox{Tr}[D(\xi)\rho]$ for the density operator $\rho$
with $D(\xi )=\exp [\xi \hat{a}^{\dagger }-\xi ^{*}\hat{a}]$.
We have found that the Wigner function of the MQS in Eq.~(\ref{MMQS})
shows the minimum negativity
at the origin of the phase space, $\eta=0$, at all times of its evolution. 
Therefore, to characterize the nonclassicality of the MQS, we focus on the value of the Wigner %%@
function $W(\eta=0)$, 
which is essentially the photon-number parity of the state [23]. 
It is given by 
\begin{equation}
W(0)=\frac{4\Big[\frac{e^{-\frac{2\alpha^2\kappa}
{A}}}{A}-\frac{4e^{-\frac{2\alpha^2(1-\kappa)}{B}}}{B}\Big]}
{\pi\Big[2-\frac{8 e^{-\frac{2\alpha^2(1+3\kappa)}{C}}}{C}\Big]}
\end{equation}
where $A = \kappa (V-1)+1$, $B = -\kappa(V-1) +(V+3)$,
$C = 3\kappa(V-1) + (3 + V)$.
When $\gamma t=0$, the minimum
negative values of the Wigner functions
of the pure and the mixed MQSs are 
$-2/\pi$ ($\approx -0.64$) regardless of $\alpha$ and $V$.

We plot the minimum negativity of the Wigner function, $W(0)$, of
pure MQSs as a function of time in Fig.~\ref{pure}, where
evolutions of pure MQSs for $\alpha=20$, $\alpha=30$ and $\alpha=50$ have been compared.
As was pointed out in Ref.~[11-13], it is obvious that
the negativity approaches zero faster with the larger amplitude $\alpha$, i.e., when the two %%@
coherent states
$|\alpha\rangle$ and $|-\alpha\rangle$
are more separated in the phase space.

\begin{figure}
\centerline{\scalebox{0.75}{\includegraphics{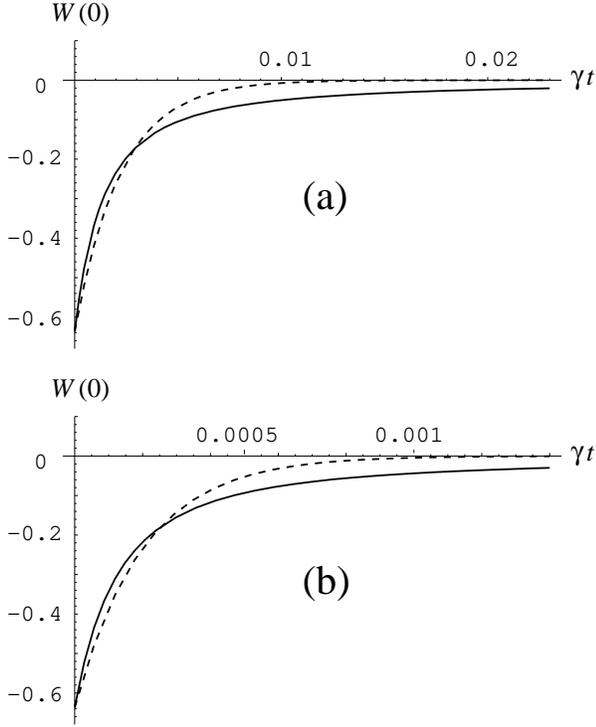}}}
\caption{
The minimum negative values of the Wigner function
of pure MQSs (dashed curves) and of highly-mixed MQSs (solid curves)
for the same separation (i.e. the same $\alpha$)
between the component states.
(a) A pure MQS with $\alpha=30$, and a highly-mixed MQS with $V=10^3$
and $\alpha=30$. The average photon number 
of the pure MQS is 900 while that of the highly-mixed MQS is $\approx 1.4\times10^3$.
(b) A pure MQS with $\alpha=100$, and 
a highly-mixed MQS with $V=10^4$ and $\alpha=100$. The average photon 
number of the pure MQS is $10^4$ while that of the highly-mixed MQS is $\approx 1.5 \times 10^4$.
In both cases, the minimum negative values of pure MQSs
approach zero faster than those of highly-mixed MQSs after a certain time.}
\label{fig-sep}
\end{figure}

We now compare the negativities of pure MQSs and of mixed MQSs under the condition that
their component states are equally separated.
In Fig.~\ref{fig-sep}(a), the pure MQS of $\alpha=30$
are compared with a mixed MQS of $V=1000$ and $\alpha=30$.
Note that the mixed MQS is highly mixed as $S(\rho)\approx 0.999$.
As shown in the figure, the negativity of the mixed MQS
reduces more rapidly at the initial stage and thus
negativity of the pure state is deeper for $t< 0.0025$.
However, Fig.~\ref{fig-sep}(a) shows that the negativity of the mixed MQS
remains deeper for $t>0.0025$ and reduces slower than that of the pure MQS.
In Fig.~\ref{fig-sep}(b), where a similar trend is observed, the pure MQS of $\alpha=100$
is compared with the mixed MQS of $V=10^4$ and $\alpha=100$, of which
degree of mixedness is $S(\rho)\approx 0.9999$.

\begin{figure}
\centerline{\scalebox{0.75}{\includegraphics{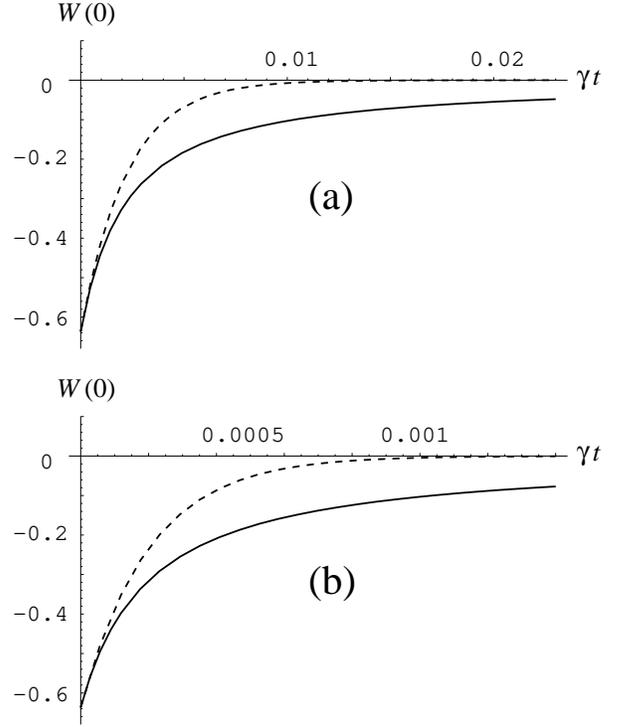}}}
\caption{
The minimum negative values for the Wigner functions
of pure MQSs (dashed curves) and highly-mixed MQSs (solid curves)
 for the same average photon numbers:
(a) a pure MQS with $\alpha=30$, and a highly-mixed MQS with $V=10^3$
and $\alpha=20$, where the average photon number of each state
is equally $\approx 900$, and (b) a pure MQS with $\alpha\approx 100$, and 
a highly-mixed MQS with $V=1.5\times 10^4$ and $\alpha=50$, where
the average photon number of each state is equally $\approx 1.0\times 10^4$.
The minimum negative values of the pure MQSs
obviously approach zero faster 
than those of the highly-mixed MQSs.
}
\label{fig-size}
\end{figure}

We also make comparison of decoherence between pure and mixed MQSs that
are equally macroscopic in view of the average photon number.
%It should be noted that for this comparison, we refer to
%sizes of physical systems.
In these cases, the component states of the mixed MQSs should be less separated
than those of the pure MQSs in the phase space.
In Fig.~\ref{fig-size}(a), 
a pure MQS with $\alpha=30$ and a highly-mixed MQS with $V=10^3$
and $\alpha=20$ are compared, where the average photon number
is ${\bar n}\approx 900$ for both states.
Here, the degree of the mixed MQS is $S(\rho)\approx 0.999$.
Figure~\ref{fig-size}(a) clearly shows that the mixed MQS
decoheres more slowly. In Fig.~\ref{fig-size}(b),
a pure MQS with $\alpha=100$ and a highly-mixed MQS ($S(\rho)\approx 0.9999$) with $V=1.5\times %%@
10^4$
and $\alpha=50$ are compared, where the average photon number is all
${\bar n}\approx 1.0\times 10^4$.
The same trend is also manifest in Fig.~\ref{fig-size}(b).

At first sight, the above results might look somewhat counter-intuitive, as one usually expects 
that the mixedness plays a negative role as degrading quantum effects. 
These findings can be accounted for by carefully rearranging terms for the mixed MQS in %%@
Eq.~(\ref{MMQS}) to give
\begin{eqnarray}
\rho=\int d^2\beta P_{\{V,\alpha\}}(\beta)|\Psi_\beta\rangle\langle\Psi_\beta|
\label{MMQS1},
\end{eqnarray}
where
\begin{eqnarray}
P_{\{V,\alpha\}}(\beta)=\frac{2}{\pi(V-1)}\frac{1-e^{-2|\beta|^2}}
{1-\frac{1}{V}e^{-\frac{2}{V}|\alpha|^2}}e^{-\frac{2}{V-1}|\beta-\alpha|^2}
\label{MMQS1co},
\end{eqnarray}
and the state $|\Psi_\beta\rangle=N_\beta\left(|\beta\rangle-|-\beta\rangle\right)$
is the normalized pure MQS ($\beta$: complex). 
Now, the mixed MQS in Eq.~(\ref{MMQS1}) may be interpreted 
as the classical mixture of the pure MQS,$|\Psi_\beta\rangle$, with $P_{\{V,\alpha\}}(\beta)$ as the %%@
weighting function. 
The quantum interference of each component state
$|\Psi_\beta\rangle$ will decay at the rate of $2\gamma|\beta|^2$, as $e^{-2\gamma|\beta|^2t}$ [11].
As the probability distribution $P_{\{V,\alpha\}}(\beta)$ is centered at $\alpha$ with the variance %%@
determined by $V$ in the phase space, the mixed state $\rho$ is composed of both of the faster and %%@
the slower decaying pure states than the single pure coherent superposition $|\Psi_\alpha\rangle$. %%@
In the long-time limit, only the slower decaying states will survive, which constitutes the basis %%@
for the overall slower decoherence of the mixed MQS in Fig.~\ref{fig-sep}. 
The difference in decoherence between the mixed MQS and the pure MQS in Fig.~\ref{fig-size} can also %%@
be similarly explained.

\begin{figure}
\centerline{\scalebox{1.2}{\includegraphics{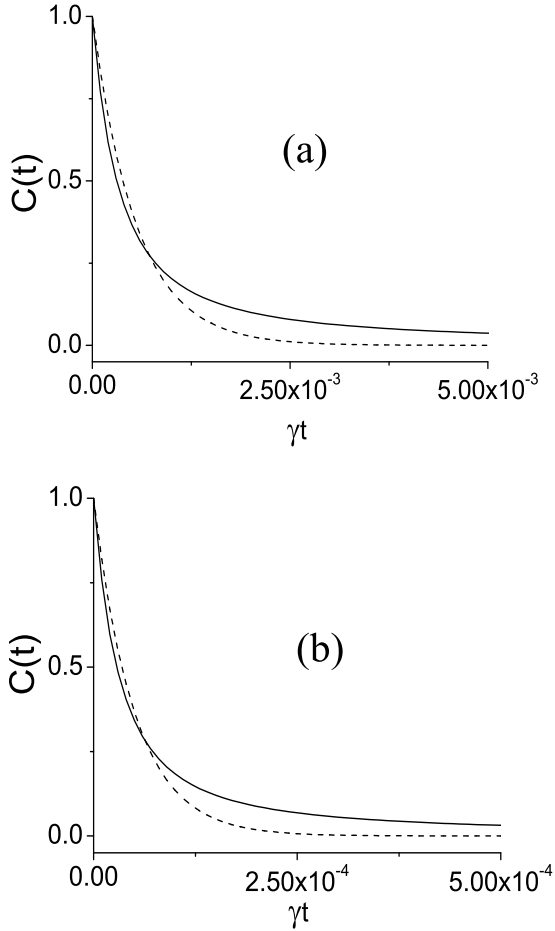}}}
\caption{The probabilistic mixture $C(t)$ in Eq.~(\ref{dec}) (solid curves)
 of exponentially decaying terms, as compared with the single decaying term 
 $e^{-2\gamma|\alpha|^2t}$ (dashed curves), for (a) $\alpha=30$, $V=10^3$, and 
(b) $\alpha=100$, $V=10^4$.}
\label{ref}
\end{figure}

To illustrate the present situation, let us consider the mixture of decaying terms %%@
$e^{-2\gamma|\beta|^2t}$ with the probability density $P_{\{V,\alpha\}}(\beta)$, that is, 
\begin{eqnarray}
C(t)&=&\int d^2\beta P_{\{V,\alpha\}}(\beta)e^{-2\gamma|\beta|^2t}\nonumber\\
&=&\frac{1}{1-\frac{1}{V}e^{-\frac{2}{V}|\alpha|^2}}\left[D(\gamma t)-D(\gamma t+1)\right],
\label{dec}
\end{eqnarray}
where
\begin{eqnarray}
D(x)\equiv\frac{1}{1+(V-1)x}e^{-\frac{2|\alpha|^2x}{1+(V-1)x}}.
\end{eqnarray}
In Fig.~\ref{ref}, we plot $C(t)$ with the same values of $\alpha$ and $V$ as the ones in %%@
Fig.~\ref{fig-sep}. 
Although $C(t)$ does not necessarily represent the same physical context as the Wigner function %%@
$W(0)$ over time, the plot reveals a similar trait: $C(t)$ decays faster in early time and slower
in long time than the single decay term $e^{-2\gamma|\alpha|^2t}$.

\section{Conclusions}

In this paper, we have compared decoherence evolutions of 
pure MQSs and mixed MQSs, and our study has revealed previously unknown aspects of
decoherence phenomena.  
In particular, we have found that a highly mixed MQS (actual state)
can show significant nonclassicality even at times when the corresponding 
pure MQS (target state) loses its quantum property almost completely. 
The mixed MQS is more macroscopic in view of energy quanta (mean photon number),
and our result thus implies that the energy size of the macroscopic system may
not be a good indicator of the decoherence rate beyond the very short time regime.
This has been further evidenced by comparing the pure and the mixed MQS having 
the same energy quanta in Fig.~\ref{fig-size}.
Note that the mean photon number ${\bar n}$ of the system decays as
${\bar n}= {\bar n}_0e^{-\gamma t}$ from the master eq.~(\ref{master-eq}),
thus it takes the same value at all times for both of the pure and the mixed
MQS in Fig.~\ref{fig-size}. 
  
Our findings also have a practical implication: In an optics experiment to generate a MQS,
although our target state is a pure MQS, the actual output state will be in most cases a 
mixed MQS due to the inevitable experimental noise added to the initial state. The nonclassicality
of a MQS, however, can be still observable in moderate times after it is exposed to an 
environment. Although the noise was modeled as a Gaussian one in this paper, 
the main result does not change for different types of noise, as long as the initial mixed MQS 
possesses as its components the pure MQS's that are more robust against decoherence than the single %%@
pure MQS.  
Decoherence properties of two-mode MQSs [24],
in relation to quantum entanglement and
quantum nonlocality, deserve further investigations.

\section*{Acknowledgments}

HJ was supported by the Australian Research Council, Queensland State Government
and JL by Korean Research Foundation Grant funded by the Korean Government (MOEHRD)
(KRF-2005-041-c00197). HN acknowledges the financial support by Qatar Foundation.

%\section*{References}


\begin{thebibliography}{99}
\bibitem{note} A more general form
of such a superposition state is $|\alpha\rangle+e^{i\varphi}|-\alpha \rangle$,
where the normalization factor is omitted and $\varphi$ is arbitrary,
while we choose $\varphi=\pi$ for convenience.
However, all the conclusions in this paper remain the same 
regardless of the choice of $\varphi$.
\end{thebibliography}
\end{document}